\documentclass[12pt]{report}

\usepackage[a4paper, margin=1in]{geometry}
\usepackage{graphicx}
\usepackage{hyperref}
\usepackage{amsmath, amssymb}
\usepackage{enumitem}
\usepackage{titlesec}
\usepackage{tocloft}
\usepackage{tikz}

\titleformat{\section}{\normalfont\large\bfseries}{\thesection.}{1em}{}
\titleformat{\subsection}{\normalfont\normalsize\bfseries}{\thesubsection.}{1em}{}

\title{Chapter 46 \\
Floods and Droughts in Asia, Europe, and America}
\author{
  Hiroshi G. Takahashi\textsuperscript{1},
  Masashi Kiguchi\textsuperscript{2},
  Shiori Sugimoto\textsuperscript{3} \\
  \\
  \textsuperscript{1}Tokyo Metropolitan University \\
  \textsuperscript{2}Institute for Future Initiatives, The University of Tokyo \\
  \textsuperscript{3}Japan Agency for Marine-Earth Science and Technology
}
\date{April 2022}

\begin{document}

\maketitle
\pagestyle{plain}

\chapter{Floods and Droughts in Asia, Europe, and America}

\tableofcontents

\begin{abstract}
This chapter introduces flood and drought through the understanding of
the water cycle. In addition to the water cycle, we consider the energy
cycle. The floods and droughts have strong regional and seasonal
characteristics. The causes of the unbalanced water conditions can occur
under the various meteorological phenomena, which have strong regional
and seasonal varieties. For an understanding of the cause of floods and
drought, we first consider the geographical characteristics of the
floods and droughts. At the same time, we focus on the spatial and
temporal time-scale of the flood or drought. Moreover, because floods
and drought can be considered as excess and shortage of water,
respectively, they are opposite. However, their spatial and temporal
scales are asymmetric.
\end{abstract}

\textbf{Keywords:} Climatic zone, seasonal and regional differences,
precipitation characteristics, temporal and spatial scales

\section{Introduction of Flood and Drought}

This section explains the flood and drought in Asia, Europa, and America
in terms of meteorological and climatological scientific viewpoints. To
understand more practical countermeasures against floods, it is also
better to see a hydrological handbook.

Generally, floods and droughts have severe impacts on society all over
the world. To understand the floods and droughts, we should understand
the concepts of "hydrological balance", which can be simply explained
by the difference between the input and output of water on a considering
area (Fig. 1), such as a watershed and a part of a watershed. When the
input of water into the considering area is larger than the output of
water, the mass of water must be increased and, in turn, can induce a
flood. On the other hand, if the input is smaller than the output, a
drought condition may occur. To understand the "hydrological balance"
in the considering area, we should understand the hydrological cycle.
The hydrological cycle has many processes, such as Ocean precipitation,
Ocean evaporation, water vapor transport, land precipitation,
evaporation, transpiration, surface flow, percolation, groundwater flow
(Fig. 2). In addition, we have to think of the balance of energy as well
as the balance of water (Fig. 2).

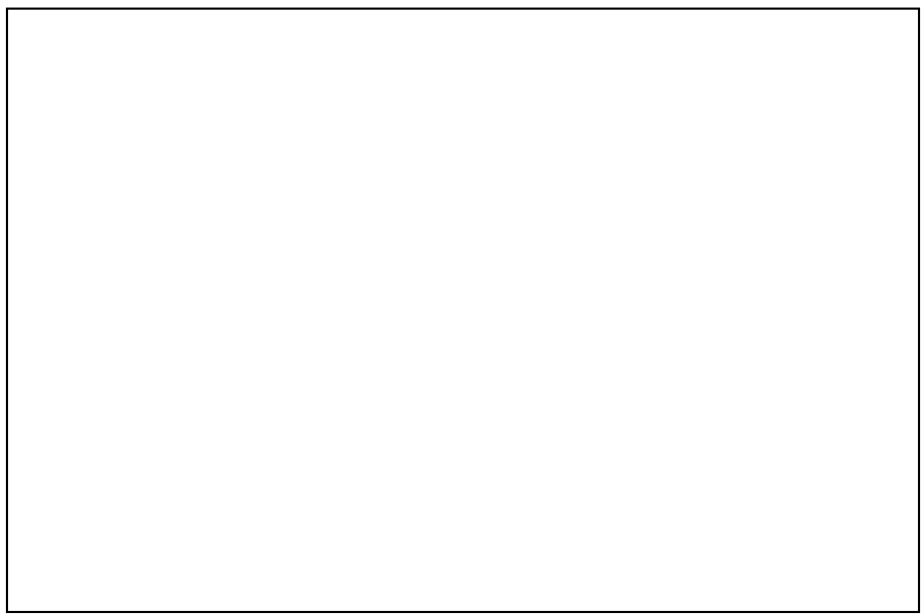
\begin{figure}[htbp]
  \centering
  \begin{tikzpicture}
    \draw[black, thick] (0,0) rectangle (12,8);
  \end{tikzpicture}
  \caption{\textit{Figure omitted due to copyright. See: Trenberth et al. 2007.} Schematic of the local atmospheric water balance. The large arrows indicate atmospheric moisture divergence, which is mostly compensated for by evapotranspiration $E$ and precipitation $P$, as changes in atmospheric moisture storage are small. At the surface $E-P$ is balanced by surface and subsurface runoff, and changes in soil moisture and groundwater. (Citation: Fig. 2 of {[}1{]} Trenberth et al.~2007)}
\end{figure}

The inputs of water are basically direct precipitation flux in the
vertical direction from the atmosphere, and surface flow and underground
water flow in the nearly horizontal direction. The outputs of water are
basically evaporation fluxes from the land or water body from the
considering area, surface flow, and groundwater flow into the out of the
considering area. However, all physical processes, such as
precipitation, surface flow, are very complicated to monitor, understand
and predict the individual components, and the quantification of the
total budget of water is still very difficult.

\begin{figure}
\centering
  \begin{tikzpicture}
    \draw[black, thick] (0,0) rectangle (12,8);
  \end{tikzpicture}
\caption{\textit{Figure omitted due to copyright. See: Trenberth et al. 2007.} The hydrological cycle. Estimates of the main water reservoirs, given in plain font in 10$^{3}$ km$^{3}$, and the flow of moisture through the system, given in slant font (10$^{3}$ km$^{3}$ yr$^{-1}$), equivalent to Eg (10$^{18}$ g) yr$^{-1}$. (Citation: Fig. 1 of {[}1{]} Trenberth et al.~2007)}
\end{figure}
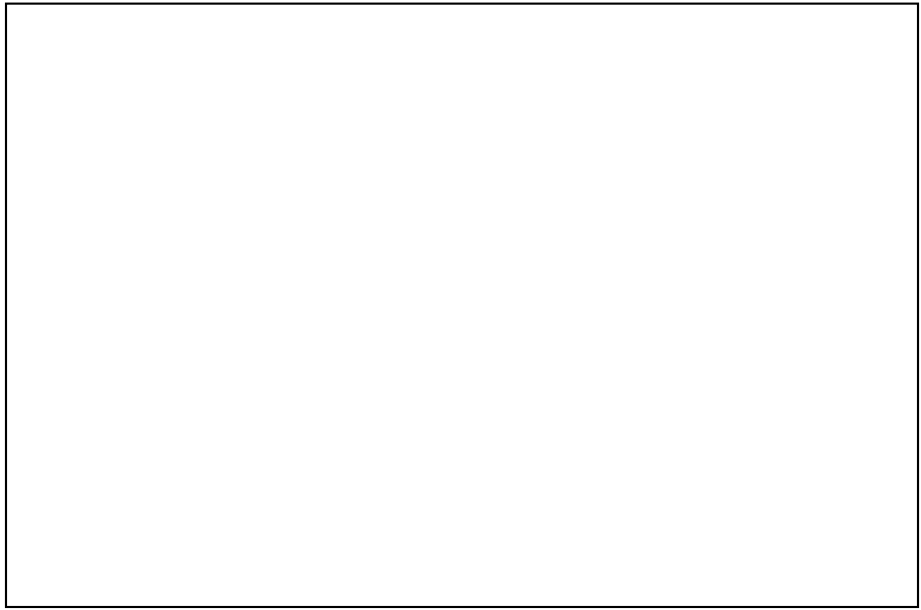

In addition to this, we should consider the temporal and spatial scales
of the physical processes over the world. Such a flash flood generally
in a shorter time scale than one day, short-time (time-scales from tens
of minutes to hours) and heavy precipitation, and surface flow may be
essential drivers. Some other processes, such as evaporation from the
surface and underground flow, can be quantitatively negligible because
the terms of evaporation and underground flow are much smaller than
those of precipitation and surface flow. However, this flash flood may
occur at the small spatial scales, which cannot induce a continental
scale flood. When we consider a continental scale flood, large-scale
atmospheric processes can be a significant driver, such as tropical
storms, mid-latitude low-pressure systems, and mid-latitude frontal
systems.

On the other hand, drought can be due to a lack of water. Over the
tropics, El Niño Southern Oscillation (ENSO) can induce a long-lasting
(more than one month) drought. Over the mid-latitude regions,
atmospheric blocking high, which generally lasts a few weeks, can induce
a continental scale drought. It is noteworthy that the spatial scales of
droughts are generally larger than those of the floods. In addition to
this, the temporal scale of droughts is longer than that of floods.
Moreover, droughts are generally associated with heatwaves.

Here, it should be noted that floods and droughts can be the opposite
phenomenon because the excess water in the considering area can induce a
flood, and the water shortage in the considering area may induce a
drought. However, they are clearly asymmetric on temporal and spatial
scales. Thus, physical phenomena can be the opposite; however, they
should not be so simple.

The concept of geography is quite essential to understanding floods and
drought. The atmospheric and meteorological characteristics have a
strong dependence on the regions. Air temperature, humidity, and winds
are quite different among the climatic regions, which is associated with
floods and droughts. For example, in the tropics, tropical storms can be
a significant factor. In the mid-latitude regions, mid-latitude
low-pressure systems and mid-latitude frontal systems should be
considered, although mesoscale precipitation systems embedded in the
large-scale atmospheric systems may be critical in some cases.

Also, land surface conditions, such as soil and vegetation conditions,
can be associated with floods and droughts, although they are also
related to the climate, as seen in the Köppen climate zone. The
coefficient of water permeability of the land surface is associated with
the output of water from the considering area. Vegetations enhance
transpiration, which also affects the water budget of the considering
area. In addition, human-induced changes in land surface conditions,
such as urbanization and cultivation, strongly influence the water
budget of the considering area. However, because the human-induced
changes of land surface conditions are a wide range of variations, we
discuss the impacts in the minimum necessary.

Moreover, small and large-scale topographies should be considered. This
is because the hydrosphere response is quite different under the
different topographic conditions, even if the same heavy precipitation
event occurs. Thus, this chapter also provides the characteristics of
flood and drought, focusing on the different responses of flood and
drought due to the regional and seasonal differences. We should
understand future regional predictions for deepening scientific
knowledge and decision-makers.

\section{Water and energy cycles for the understanding of flood and drought}

Flood and drought can occur due to an imbalance of water and energy over
a target region. The spatial and temporal scales of the imbalances are
dependent on the flood and drought events. The imbalance can occur under
the anomalous atmospheric circulations, such as short-duration heavy
precipitation, abundant seasonal precipitation, a development of the
drier weather than usual due to a blocking high, and so on. The
anomalous atmospheric circulations are associated with the regional
characteristics of climatology.

The imbalance of surface water directly explains the floods and
droughts. In addition to the imbalance of surface water conditions, the
imbalance of surface energy is closely related to the imbalance of
surface water conditions. This is because the term of the latent heat
flux in the surface water budget is tied with the redistribution of the
surface available energy, which is derived from the shortwave radiation
(solar radiation) and longwave radiation (atmospheric radiation).

Specifically, under drier surface water conditions, more energy is used
for raising the surface air temperature. In this situation, less energy
is used for the vaporization of surface water, which in turn relatively
decreases atmospheric water vapor. In the actual cases, the
redistribution of the surface energy is related to the surface air
temperature in the target conditions. There is a temperature dependency
of the redistribution of the surface energy. More specifically, in
higher temperature conditions, more energy is used for vaporization.
This can also be explained by using the Bowen ratio in some cases. Bowen
ratio (Bo) is defined as the non-dimensional ratio of sensible heat
divided by latent heat (Bo = H / L\textsubscript{e}E). Here, H is
sensible heat, which is used for the rising temperature,
L\textsubscript{e} is latent heat required for water evaporation, and E
is vaporized water. A larger (smaller) Bo value indicates that H is
relatively larger (smaller) than L\textsubscript{e}E, which implies that
the land surface condition is dry (wet). Thus, the Bowen ratio is a
useful value to understand the surface water conditions.

Feedback mechanisms are also very important to understand floods and
droughts. The specifics of the feedback process will be explained in the
following parts. As a result of the redistribution of the surface water
and energy, atmospheric conditions and circulations are modified on the
regional scales, particularly in the downstream regions. The changed
atmospheric conditions can modify the water and energy budget in the
target regions. This can be more significant in the case of droughts.
For example, when some trigger causes a rising surface air temperature,
additional rising surface air temperature can occur. Warmer surface air
temperature condition accelerates evaporation, and surface water
condition becomes drier than cooler atmospheric conditions. The drier
surface can use the surface energy for the rising surface air
temperature. This series of processes can be repeated as positive
feedback, which in turn induces drought conditions. To quit from this
loop, another external forcing, such as a low-pressure system, is
necessary. However, a high-pressure system, such as a mid-latitude
blocking high, may block the intrusion of the water supply, which
enhances the warm-dry feedback loop.

For understanding the flood and drought, imbalances of water and energy
should be considered simultaneously. In addition, it should be
considered how the atmospheric processes can accelerate or deescalate
the imbalance conditions at multiple spatial and temporal scales.

\section{Characteristics of flood}

A flood is one of the natural disasters and a major problem in a wide
area of the world. In 2020, EM-DAT (Emergency Events Database) reported
389 natural disasters killing 15,080 people, affecting 98.4 million
others, and costing 171.3 million US\$ ({[}2{]} EM-DAT, 2021). Compared
with the previous two decades (2000-2019), there were 26\% more storms
than the annual average of 102 events, 23\% more floods than the annual
average of 163 events, and 18\% more floods than the annual average of
5,233 deaths. These impacts of the events were not equality of location
where it occurred. The impacts of floods were heavily throughout Africa
and Asia. In Africa, floods affected 7 million people and caused 1,273
deaths, the highest figure since 2006. In South and Southeast Asia,
monsoon flooding associated with landslides frequently affected some
million people. China also has been damaged by significant flooding, and
economic loss increases due to recent economic growth.

Flooding is divided into six categories by the United States Geological
Survey (USGS) ({[}3{]} Burt, 2004). These six categories are as follows.

\begin{itemize}
  \item \textbf{\textit{Regional floods}} -- where a river overflows its banks on a large
  scale, flooding entire regions.

  \item \textbf{\textit{Flash floods}} -- when an intense local rainfall causes a stream or
  river to suddenly flood a small area, usually the local watershed of
  that river.

  \item \textbf{\textit{Ice-jam floods}} -- when melting ice-floes dam a portion of a
  river, normally in the spring, causing it to flood upstream from the ice
  blockade.

  \item \textbf{\textit{Storm-surge floods}} -- when a sudden rise in the sea level
  inundates coastal areas, usually during intense typhoons prior to the
  landfall of the eye. Also happens on a lesser scale from intense storms
  of a non-tropical nature.

  \item \textbf{\textit{Dam failure floods}} -- often the result of faulty engineering.

  \item \textbf{\textit{Mudflow flood}} -- when the ash from a volcanic eruption washes into rivers and creates mudflow-like conditions.
\end{itemize}

Of the above, dam failure flood is out of scope in this chapter. Most
dangerous floods are those caused by storm surges during Typhoons (e.g.,
Typhoon Haiyan in November 2013) and flash floods caused by intense
local rainfalls (e.g., Uttarakhand in North India in June 2013). The
typhoon forecast has recently improved more, while the forecast of the
pass way of typhoons is still difficult. Storm surge by typhoons is
strongly affected by pass way because the relationship between the wind
direction, and the coastal line, and the 3-dimensional shape of the gulf
affect the scale of the storm surge strongly. Moreover, the simulation
for storm surges has large uncertainty. On the other hand, no one knows
of any possible floods as the cause of flash floods since it is not
always necessary to have intense rainfall in the area where the flood
occurs. Moreover, the lead time for evacuation from the flash flood is
not enough due to the rapid impact on that.

\subsection{Regional floods}

Regional floods are more destructive than, but usually not as deadly as,
flash floods. Rainfall has fallen continuously in the period from a few
days to a few weeks, or seasonal rainfall amounts are very abundant in a
wide region. As a result of the abundant precipitation, flooding occurs
slowly rather than flash floods. Continuous rainfalls are provided by
the strengthened trough or front and by the active monsoon rather than
the normal on a large scale. Regional floods cause widespread property
damage and can result in large death tolls. Recently, the typical
regional flood occurred in the Yangtze River basin, located in southern
China in 2020, and resulted in around 400 deaths and property damage of
21.8 billion US\$ ({[}2{]} EM-DAT, 2021). Heavy rainfalls caused by the
Meiyu (regional rainy season) led to floods severely affecting large
areas of southern China.

\subsection{Flash floods}

Flash floods are rapid flooding of lower places below the surrounding
area (i.e., valleys, including rivers). It might be caused by heavy
rainfall associated with storms (i.e., tropical cyclones, thunderstorms,
melting water from snow or ice). Flash floods may also occur by
collapsing natural debris or ice dams. The intensity of rainfall, the
location and distribution of the rainfall, the land use and topography,
vegetation types and growth/density, soil type, and soil water-content
all determine just how quickly the flash floods may occur and influence
where it may occur ({[}4{]} National Weather Service, 2021). For
instance, flash floods occurred in Jakarta (capital of Indonesia) and
its metropolitan area on January 1, 2020, due to the heavy overnight
rainfall, causing the overflow of the rivers. The death toll is over 60.

\subsection{Ice-jam floods}

An ice jam occurs when chunks of ice clump together to block the flow of
a river. It can cause flooding near the river. These are caused by
snow/ice melting water in the spring season, and warm temperature and
rainfall in this season cause snow and ice to melt rapidly. This
additional water makes the river discharge increase, and this stream
carries an amount of ice chunks downstream. Some ice chunks may be stuck
in a narrow width of the river, then these ice chunks form an ice jam,
which blocks the river flow. This flooding often occurs in Siberia,
northern Europe, and the northeastern American continent. Ice-jam flood
in Ob River in Siberia commonly occurs and affects the infrastructures
related to human activities ({[}5{]} Papa et al., 2007).

\subsection{Storm-surge floods}

Storm surge is a rise of water generated by a storm over and above the
predicted astronomical tides. It should not be confused with storm tide,
which is defined as the water level rise due to the combination of storm
surge and the astronomical tide ({[}6{]} NOAA, 2021). It is known that
storm surge is produced by water being pushed toward the shore by the
force of the winds moving cyclonically around the storm. So, the storm's
air pressure and traveling direction are important elements for storm
surge. Along the coast, storm surge is often a huge impact on human life
and property by tropical cyclones (typhoons, cyclones, and hurricanes
developed in the northwestern Pacific Ocean, Indian Ocean and the
southern Pacific Ocean, and the northeastern Pacific Ocean and the
Atlantic Ocean, respectively). According to EM-DAT, the "Bhola"
cyclone in 1970 had attacked East Pakistan (present-day Bangladesh) and
West Bengal state in India and caused the storm surge flooding. At least
500,000 people were killed by mainly storm surge floods around the lower
area, including islands of the Ganges Delta.

\subsection{Mudflow flood}

Mudflow is a geological phenomenon in which water includes masses of
soil and rock, rush down the slopes, and a very rapid to the extremely
rapid surging flow of saturated plastic soil in a steep channel,
involving significantly greater water content relative to the source
material ({[}7{]} Hungr et al., 2014). Compared with rock avalanche and
debris flows, the size constituent particles are small. Triggers of
mudflow might be heavy rainfall, snowmelt, volcanic eruption,
earthquake, and so on. The largest historic mudflow occurred by the
eruption of Mount St.~Helens, a volcano in the state of Washington, the
USA, in 1980. The volume of material displaced was 2.8
km\textsuperscript{3}.

\subsection{Historical record of flood}

Here, the historic recorded floods are introduced based on EM-DAT and
{[}8{]} Gunn (2008). Record of extreme meteorological phenomena is
important for recognition of what would occur in the future. EM-DAT set
the criteria of disaster; 1) 10 or more people dead, 2) 100 or more
people affected, 3) The declaration of a state of emergency, and 4) A
call for international assistance. Following records follow these
criteria.

Worst historic flood occurred by the abundant rainfall during the rainy
season (July-August) in Yangtze and Huai river basins in 1931. Death
toll is estimated of 3.7 to 4.0 million ({[}9{]} Coutney, 2018).~

Second worst occurred by the breaking dike of the Yellow river in
September 1887. Death toll is at least 0.9 million, highest estimated
death toll is 2 million (e.g., {[}8{]} Gunn, 2008). Flood water spread
quickly throughout the North China Plain, covering an estimated
130,000km\textsuperscript{2} ({[}8{]} Gunn, 2008).~

Third worst one occurred by the storm-surge due to the 1970 Bhola
cyclone that attack East Pakistan (present-day Bangladesh) and West
Bengal state in India on November 11, 1970. Death toll is at least 0.5
million due to mainly storm-surge for the lower Ganges delta including
small island. The surge height was about 10 meters ({[}10{]} Kabir et
al., 2006).

Forth worst one occurred by the storm-surge due to the 1839 tropical
cyclone that attack southeast India on November 25, 1839. Coringa
located in Andhra Pradesh State of India suffered huge damage by strong
wind and a large storm-surge (about 12 meters), and death toll is around
0.3 million. Coringa has destroyed completely, then people moved to far
inland.

Fifth worst one occurred by the rainfall during rainy season along the
Yangtze River basin in July 1935. Death toll is around 145,000 (EM-DAT).
China has not recovered from 1931 Yangtze food. As a result of this, the
remaining flood relief infrastructure which included damage reservoirs
and flood water channels were soon overwhelmed.~

Top 5 worst disaster of floods are introduced so far. Here it is shortly
introduced that we exclude the floods from the historical floods list,
even though the death toll is rather than 0.2 million. Reason of
excluding is those floods occurred by human activities such as war and
dam failure.~

First one is the 1938 Yellow River flood, named as "Huayuankou
embankment breach incident", created by the Nationalist Government in
central Chida during the early stage of the second Sino-Japanese War to
halt the rapid advance of Japanese forces. Death toll (not including
military loss) is around 0.4 -- 0.9 million ({[}11{]} Muscolino, 2014).
Until the dike was rebuilt in 1947, the mainstream of the Yellow River
flow down to Huai River, and the farmers on that area suffered frequent
flood and famine.

Second one is the 1975 Banqiao Dam failure in August under the influence
of Typhoon Nina. Death toll is 26,000 -- 240,000 with huge inundation
area (about 12,000km\textsuperscript{2}) ({[}12{]} Xu et al., 2008). Two
major dams, including Banqiao Dam, two medium dams, and small 58 dams
failed from overtopping in the storm.

\section{Characteristics of drought}

Generally, three types of droughts have been defined. The following
subsections provide general categories of droughts. However, the three
types of droughts are not always independent. The classification is
based on the areas, sections, or society of the influence. In addition,
even when there are similar meteorological and climatological forcing
associated with the droughts, the response of climates varies by region.

The conditions for a drought of occurrence vary from region to region.
For example, the forcing of a monthly precipitation anomaly from the
normal climate conditions -50 mm is significant over semi-arid regions.
In contrast, it has a small impact on wet tropical regions. In
meteorology and climatology, we frequently use the anomaly from the
normal year or ratio to the normal year because the spatial distribution
of a meteorological variable, particularly precipitation, shows strong
non-uniformity. Note that the climatological values are generally
calculated to be averaged meteorological variables, such as
precipitation, surface air temperature, for a 30-year period. As a side
note, in a world undergoing climate change, the determination of
climatological values can also be a major issue.

\subsection{Meteorological drought}

A prolonged less precipitation condition than the normal condition over
a target region is generally understood as meteorological drought or
simply a drought. The meteorological drought condition is simply
understood as an imbalanced water condition at the surface. Thus, an
occurrence of a meteorological drought can be found from precipitation
observation if we have precipitation observations over the target
region. However, understanding past meteorological droughts, for
example, is often hampered by the limited availability of precipitation
observations.

In general, we can explain that a time scale of an occurrence of the
meteorological drought is longer than a few weeks. However, it should be
noteworthy that it highly depends on the climate conditions of the
target region. At least, most time and spatial scales of drought
conditions are longer and broader than their flood conditions when we
consider the same target region. Many cases of drought are accompanied
by heatwaves, and the two may not always be distinguishable.

\subsection{Hydrological drought}

Due to a meteorological drought or a long-term continuity of
meteorological drought conditions, hydrological conditions become drier
than a normal condition of a target region, such as surface flow,
groundwater levels, and soil moisture content. Because the time constant
of the hydrosphere is generally longer than that of the atmosphere,
hydrological droughts take longer to occur than meteorological droughts.

\subsection{Agricultural drought}

Meteorological drought and/or hydrological drought results in adverse
effects on the crop and vegetation. Under the higher surface air
temperature and/or drier soil moisture conditions, the plants are not
able to use the water they need, and they are not able to do enough
transpiration. The meteorological drought and hydrological drought
conditions are associated with the prolonged fewer precipitation
conditions. The less active vegetation conditions may also affect the
atmosphere in the near future if precipitation occurs later due to the
decrease of the function of the transpiration of the damaged plants.

\subsection{A feedback mechanism between meteorological, hydrological, and agricultural
droughts}

Drought events are mostly related to meteorological droughts,
hydrological droughts, agricultural droughts, and heatwaves. In
addition, there is a feedback mechanism between the meteorological,
hydrological, and agricultural droughts. When the soil moisture content
decreases than normal conditions, the drier surface condition can, in
turn, increase the surface air temperature through the changes in the
surface heat and moisture fluxes. As a result, the drier surface
conditions with the higher surface air temperature, which are sometimes
heatwave conditions, can result in further drought conditions with much
higher surface air temperature. This feedback is crucial in drought. To
escape from the feedback process associated with droughts, external
forces such as precipitation events are necessary.

\subsection{Basic knowledge of drought indices of droughts}

This subsection explains the basic knowledge behind the drought indices.
Useful indices are introduced in the following subsections based on the
previous review ({[}13{]} Seneviratne et al., 2012). For a scientific
understanding of the droughts, the following fundamental meteorological
values can deeply understand the detailed physical processes and
temporal evolution of the climate conditions related to the droughts.

To understand the drought conditions over the world, many indices can be
used ({[}13{]} Seneviratne et al., 2012). Some indicators are simply
understandable and easy to monitor the meteorological conditions if
real-time or quasi-real-time monitoring can be provided over the world.
Currently, satellite observations of the surface of the Earth can
provide global information of surface conditions.

Drought can be characterized as the surface air temperature conditions
and surface water conditions. In general, a drought condition is one in
which the ground surface temperature is high, or the ground surface is
dry.

During a drought period, surface air temperature is generally higher
than normal conditions. An anomaly from the climatological surface air
temperature, for example, January surface air temperature, is much
higher, which indicates a possibility of drought. However, it is
dependent on the climatological features. Thus, surface air temperature
can be an indicator of drought conditions. However, it is not easy to
determine the surface air temperature from satellite observation. To
estimate the spatial distribution of the surface air temperature from
the satellite observations, it is necessary to be fitted with some kind
of numerical model (physical or statistical).

As an indicator of the surface water condition, the surface temperature,
which is the skin temperature of the surface of the Earth, is available.
It is important to note that "\textit{surface temperature"}
and"\textit{surface air temperature"} are two completely different
variables, although the surface temperature looks similar to the surface
air temperature. Surface temperature is more largely affected by surface
water conditions than the surface air temperature. When the water in a
near-surface soil layer is small, the surface temperature increases
rapidly because of the surface energy and water balance ({[}14{]}
Seneviratne, S. I., et al.~2010). When the energy is supplied to the
surface as downward solar radiation (In meteorology and climatology, and
related sciences, visible radiation is called shortwave radiation, here,
this is a contrast with infrared radiation, i.e., longwave radiation.
This is because the wavelength of visible radiation is shorter than the
wavelength of infrared radiation.), the energy should be redistributed
into the energy for the increase in the surface temperature and surface
water evaporation. When surface water is smaller, more available energy
at the surface is used to increase in surface temperature, which in turn
increases the surface air temperature through the atmospheric turbulence
in the atmospheric boundary layer. Thus, the surface temperature can be
an indicator of the surface water condition. However, the response of
surface temperature to the surface water condition is not simple. Under
the condition of the shortage of surface water, the surface temperature
rapidly increases. Under the abundant surface water conditions, the
increase in surface temperature is suppressed due to evaporative
cooling.

As another surface water condition, surface wetness has been observed
from the satellite observations. For example, using a passive microwave
imager on satellites (e.g., Advanced Microwave Scanning Radiometer 2
(AMSR2)/Global Change Observation Mission - Water Satellite 1(GCOM-W1))
surface water conditions can be obtained around the world. This type of
observation is relatively new, which has been available since the end of
the 20th century. The surface wetness is an indicator of the drought
condition. As well as the surface temperature, the climatological
condition in surface wetness is geographically different. Thus, we
should also use anomalies of surface wetness for monitoring drought
conditions.

It is well known that precipitation is a good indicator of drought
conditions. Even today, rain gauges play an essential role, but there
are many areas where rain gauges are not available, and there are many
aspects of forecasting and monitoring that cannot be covered by rain
gauges alone. However, direct observation of the spatial distribution of
the precipitation is very limited. For example, the core satellites of
the Tropical Rainfall Measuring Mission (TRMM) and the Global
Precipitation Measurement (GPM) have radar observations from space.
However, high-latitude regions are out of the measurements. To estimate
the precipitation distribution, cloud information has been used for many
years as an alternative to precipitation, although it is difficult to
monitor precipitation quantitatively from cloud information. Currently,
several datasets of precipitation are distributed, which will be
introduced in the following section. Also, other drought-related indices
are introduced below.

\subsection{Drought indices}

While the surface temperature, soil wetness, and precipitation presented
in the previous section are themselves useful indicators of drought, a
variety of other indicators are also used. Other drought indices have
been summarized in the review {[}13{]} (Seneviratne et al., (2012) of
\textit{"Box 3-3 The Definition of Drought"}. Basically, they introduced
three types of drought indices, indices calculated from only
precipitation dataset, ones calculated from precipitation and potential
evaporation dataset, and ones calculated from more complicated numerical
models. For practical use purposes, some simple drought indices are
introduced here.

The first index is the consecutive dry days (CDD). This CDD index is
quite simple to count the consecutive days that have no precipitation.
This CDD index can be calculated from the daily precipitation of
observation and a climate simulation. CDD is quite simple but reasonably
practical for the analysis of the drought. In the context of climate
change, the CDD may change in the future. Because the long-lasting dry
days can induce a drought, the longer CDD increases the risk and impact
of the drought. Generally, the drought tends to be severe due to the
long-lasting dry conditions because soil conditions can close to the
shortage of soil moisture for evaporation. However, it may not be useful
for dry regions because dry days continue for several ten days, even in
normal climate conditions.

Also, soil moisture anomalies (SMA) are useful. SMA is easy to
understand the condition of soil moisture. If the SMA can be monitored,
the shortage of soil moisture, which induces the drought condition, can
be found. The SMA is also quite useful for analyzing the climate model
simulations because most climate models coupled with land surface models
and the soil moisture values are outputted. What time-scale SMA to
analyze depends on the climate of the region, but SMA on time scales
longer than a week is often examined. However, there is a problem with
the limited period of SMA data for analysis of past soil moisture
observation data. Compared with the precipitation or temperature data,
the soil moisture data has been limited. In addition, the observation
data before the satellite observation era are spatially limited.
Generally, soil moisture conditions are quite significant heterogeneity
in space and depth. Thus, the SMA is useful but is not available
depending on the research purpose.

As a drought index calculated from precipitation and potential
evaporation datasets, the Palmer Drought Severity Index (PDSI) ({[}15{]}
Palmer, 1965), which measures the difference of moisture balance from
normal climate conditions using a simple water balance model (e.g.,
{[}16{]} Dai, 2011), is well known and commonly used. However,
estimating the potential evaporation is not so simple because it is
difficult to observe. In addition, the target time-scale of PDSI can be
more than one year. Thus, PDSI is used for long-term droughts, such as a
multiyear drought.

In addition, many researchers have proposed and studied many drought
indices, focusing on regional drought events. However, the usefulness of
the drought indices should be considered over the target regions. As the
CDD may not be useful over the dry climate regions, individual indices
have advantages and disadvantages, which should be related to the
regional climate conditions.

\section{Precipitation datasets}

This subsection introduces the current precipitation datasets in terms
of the characteristics of temporal and spatial coverages. To monitor and
understand floods and droughts, these precipitation datasets are
essential. Without the observational datasets, we cannot monitor,
understand and discuss the current and past statuses of the flood and
drought. Thus, the precipitation datasets are the most significant
resources for flood-and-drought predictions and research.

In addition, precipitation observations have evolved in recent decades
from direct observation by the offline rain-gauge networks to the
near-online remote sensing (radar, microwave imager, infrared imager,
and so on) precipitation or their mixtures. The characteristics of
individual precipitation datasets are different, including the
uncertainty of the dataset.

As shown in below, various kinds of precipitation datasets are
available. If the target is flash flooding associated with heavy
rainfall events on small spatial scales and short time scales, products
based on precipitation radar and recent satellite data are suitable. On
the other hand, if the target is a long-term drought or a long-term
variability of droughts, it is necessary to use long-term homogeneous
precipitation data, although the spatial resolution does not
necessarily have to be fine. In general, the shorter time-scale and
spatially higher-resolution precipitation datasets are often not
suitable for analysis of long-term variation, as it often does not give
much consideration to long-term homogeneity of data quality. On the
other hand, long-term data often have a coarse resolution in space and
time and are not suitable for the analysis of phenomena with short time
scales and small spatial scales, such as flash floods and short-duration
heavy rainfall. Although a large number of precipitation data exist,
appropriate precipitation data should be selected according to the
purpose.

\subsection{Long-term land precipitation data from rain-gauge precipitation and their merged archives}

The classical and long-term precipitation datasets are based on the
direct observations by a rain-gauge itself, rain-gauge network, or their
archives (e.g., Global Precipitation Climatology Centre (GPCC), {[}17{]}
Schneider et al.~2008). Many researchers and technicians have devoted
themselves to unifying or quality-check the observed data. The primary
purpose of the quality checks is to obtain possibly homogeneous datasets
in time and space. In terms of a time axis, the qualities are quite
important to understand climate change in the target region. If the
observational point can be changed, the homogeneity cannot be preserved.
In terms of a spatial axis, the qualities are quite important to compare
the data between the different observations. If the qualities of the
precipitation observations are quite similar, we can use the data for
the spatial analysis, which is associated with the construction of
rainfall networks.

\subsection{High-resolution land precipitation data from precipitation-radar}

Due to the development of the precipitation radar technique,
precipitation-radar networks have been constructed all over the world.
The precipitation-radar networks are often operated by the national
meteorological agency. Compared with the rain-gauge, the number of
meteorological stations can be reduced. However, the observation
coverage is closely related to the topography. For example, a valley
region cannot be observed from the outside of the valley. There may be
other technical problems; temporally and spatially high-resolution
precipitation data can be obtained from a precipitation-radar network.
Currently, the high-resolution precipitation from the radar can be
obtained near-real-time, which is used for flood predictions.

\subsection{Long-term global or quasi-global precipitation from rain-gauge and satellites}

Using many satellite observations, many precipitation products have been
developed in a few decades (e.g., Global Precipitation Climatology
Project (GPCP); {[}18{]} Adler et al.~2017). It started with infrared
radiometer observations from polar and geostationary orbit satellites.
Using the infrared radiometer, we can understand the cloud activity,
which is basically related to the precipitation activity. However, the
infrared radiometer can only detect cloud-top information from space. To
improve algorithms of the precipitation estimation, other sensors were
gradually used for the estimation of precipitation. For example, a
microwave imager can detect some cloud and water vapor information; many
microwave imagers have been used for the algorithms of precipitation
estimation. In the development of the algorithms for precipitation
estimation, grand validation is very important. To improve the
estimation, ground-based observations are used. Currently, the
high-resolution precipitation from the satellites can be obtained
near-real-time, which is used for flood predictions.

\subsection{High-resolution precipitation derived from numerical models}

When we use all the satellite observations and radar observations, we
cannot predict future precipitation even in a few minutes. To predict
future precipitation activity, we use many kinds of numerical models.
For example, for precipitation forecasts shorter than one hour, we can
use simple advection models, physical numerical weather models, and
other statistical models. Currently, a statistical model can be
developed by an artificial intelligence (AI) technique.

To predict near future precipitation in a few hours, numerical
predictions can be conducted using well-initialized conditions. For the
initial values of numerical predictions, real-time or near real-time
observations are used. More accurate predictions can be made by using
initial values that sufficiently reflect the current atmospheric
conditions obtained from real-time observations.

While there are a lot of valuable datasets, there are also data that are
used despite their unreliability and should be used with caution. In the
last two decades, atmospheric reanalysis products have been developed
(e.g., {[}19{]} Kalney et al.~1996, {[}20{]} Kobayashi et al.~2015)
because of the progress of the assimilation technique and studies with
the rapid increase of computational resources.

The well-formatted atmospheric analyses are very useful, which are very
similar to the gridded precipitation datasets. It covers the entire
globe in more space and time than the data in the gridded precipitation
datasets and provides data as if it were available even where there are
no past observations. Generally, the gridded precipitation dataset is
derived from the past observations, which were interpolated but not
extrapolated. Generally, we cannot use the extrapolated precipitation
data.

However, when we use the atmospheric reanalysis, we should be careful
which meteorological variable is reliable or not. Although the
reanalysis has precipitation values, the precipitation values are very
uncertain because precipitation values have not been assimilated, i.e.,
it is the output of the simulation. There is another reason why the
precipitation values in reanalysis datasets are unreliable. However,
because the explanation is beyond the level of this paper, we briefly
explain that unreliability is associated with the uncertainty in the
physical model. To begin with, generally, the meteorological element of
precipitation cannot be considered to be suitable for assimilation.
However, the uncertainties can have been gradually reduced with
improvement of the assimilation technique, as the improvement of
assimilation of the source of the precipitation, namely water vapor.
Nevertheless, the precipitation values in the reanalysis dataset cannot
be used as observational data.

In brief, the atmospheric reanalysis is a kind of climate model output,
which is assimilated with a lot of observations over the globe. If no
observations are assimilated, the reproducibility of the atmospheric
reanalysis is quite low. Atmospheric structure and flows are not
assimilated with the observational data over the not observation area.
Nevertheless, because uncertain atmospheric reanalysis datasets are
available, the data is often used without understanding the nature of
the data, mistakenly believing it to be accurate. Thus, we should be
careful to use the datasets.

\section{Various scales of flood and drought}

Time-scales of flood and drought have a very wide spectrum, which ranges
from a few dozen minutes, a few hours, a few days, a few months, and
possibly a few years. Generally, the time-scales of floods and droughts
have strong regional dependencies. Also, because meteorological
characteristics can change seasonally, floods and drought also have
strong seasonal dependencies. It should be noted that the time-scales of
floods and droughts are not symmetric, even if we focus on a specific
region. The opposite of flood is not just drought. Thus, it should be
better to discuss them separately.

\subsection{Various spatial scales of flood}

This subsection focuses on floods. In addition to the time-scales,
spatial scales of floods are quite important. The range of spatial
scales is across a sub-kilometer scale, a few kilometers scale, a
hundred-kilometer scale, and a continental scale. Thus, the spatial
scales of the floods are also various.

Here, the "spatial scale of floods" is closely associated with the
"temporal scale of floods". Mostly, when the spatial scale of a flood
is small, the time scale of the flood is small and vice versa. This is
very similar to the basic meteorological knowledge of scales between
time and space. Thus, for understanding the temporal and spatial scales
of floods, it is quite natural that it should be simultaneously focused
on both temporal and spatial scales and the target meteorological
phenomenon. The target meteorological phenomenon may have the same order
of magnitude of spatial and temporal scales of the flood.

\subsection{Regional characteristics of floods and the relationship between the orographic factors}

The regional characteristics of floods are basically associated with the
meteorological characteristics of the target regions. However, if the
same heavy precipitation occurs over similar tropical regions due to a
tropical cyclone, the hydrological responses to the same heavy
precipitation have a strong regional dependency. As an example, we will
introduce the different regional responses to the tropical cyclone
precipitation over the Southeast Asian monsoon regions.

Thailand experienced a severe flood in 2011 (Detailed processes are
explained in Section 46.7.3). Similar types of floods can occur over
Bangladesh and part of northern India because the topographic and soil
conditions are similar. On the other hand, the meteorological disasters
in Vietnam and the Philippines are well-known as short-time floods,
although their rainfall variability is similarly associated with
tropical cyclones.

These differences can be explained by the orographic features. When we
look at Asian countries, the orographic characteristics are quite
different. Some counties have steep orography, other countries have a
broad plain, which is quite gentle orography. In some island countries
with steep orography, short-term flash floods occur at relatively
shorter time-scales, such as the order of hours. Considering the time
for the receding of the flooded water, it takes several days. However,
over the flat plains of continents, seasonal floods occur as the
seasonal march of climate or had occurred. The time-scale of the floods
can be a few months commonly. Some areas have a time-scale intermediate
between flash floods and seasonal floods.

Therefore, if the same precipitation occurs over the various orographic
regions, the precipitation responses are pretty varied, which can
explain the regional differences in floods.

\subsection{Drought and heatwaves and related to air pollutions}

Compared with floods, droughts and heatwaves basically have synoptic or
large spatial scales. Thus, the spatial scale of a drought and a
heatwave is much larger than the floods over the same target regions.
When a heatwave occurs over India, most parts of the country experience
the heatwave at the same time. The spatial scales of the droughts and
heatwaves are larger than the spatial scale of the country. When a
heatwave occurs over western Europe, the heatwave covers multiple
countries.

Thus, the spatial scales of the drought and heatwaves are larger than
that of the floods. In addition to the spatial scales, the temporal
scales range from a few days, a several-day period, to a few weeks,
which are longer than that of floods. In some worst cases, the dry
period continues across the season, which is associated with not only
drought or heatwaves but also large agricultural drought and a
large-scale wildfire. In the recent example, 2019--20 Australian
bushfire season (Fig. 3), California wildfire season (see also in
Section 7.5), and worldwide tropical rain forest fires are huge
drought-related disasters, which also provide a large amount of polluted
air, namely aerosols and gases. In the 2019--20 Australian bushfire
season (Fig. 3), it was clearly found that high aerosol optical
thickness (AOT) and negative direct aerosol radiative forcing (ARF) on
and over the downstream regions which was induced by the biomass burning
over Australia. The biomass burning can be captured by high surface
albedo (SA). It is noteworthy that the occurrence of bushfires is
strongly associated with the drought in Australia. Thus, drought and
heatwaves are also associated with air pollution and climate changes
(see also the latter parts of Section 7.1).

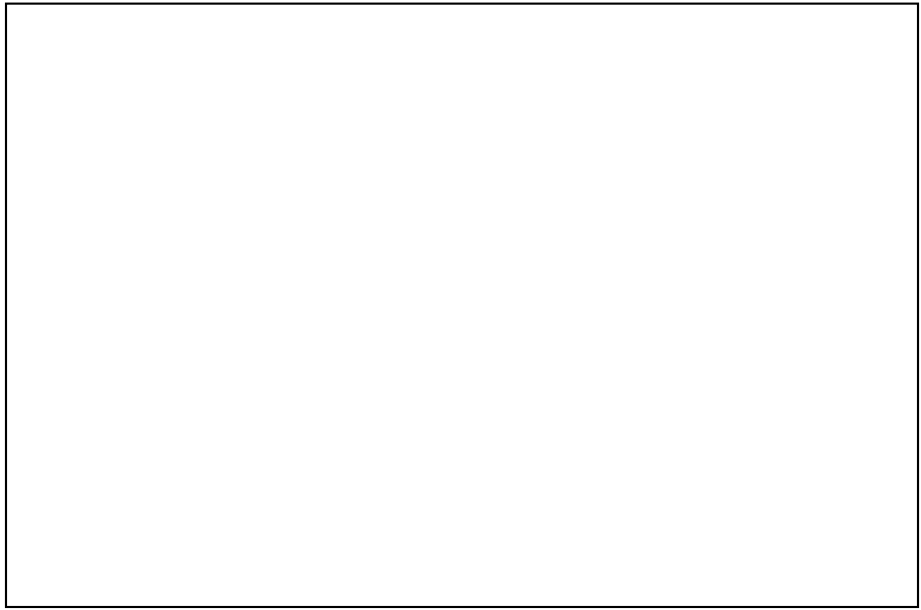
\begin{figure}
\centering
  \begin{tikzpicture}
    \draw[black, thick] (0,0) rectangle (12,8);
  \end{tikzpicture}
\caption{\textit{Figure omitted due to copyright. See: Chang et al. 2021.} The atmospheric path of biomass burning-aerosols emitted during
the 'Black Summer' from Australia across the Pacific Ocean; aerosol optical thickness (AOT), surface albedo (SA), direct aerosol radiative forcing (ARF) from 1st to 8th of
January, 2020. (Citation: Fig. 4 of {[}21{]} Chang \textit{et al.} 2021)}
\end{figure}

\section{Past floods and droughts and associated meteorological phenomena}

\subsection{Global features}

Many studies have reported the long-term changes in floods and droughts.
However, the long-term changes in floods are very difficult to evaluate
because people have developed countries against heavy precipitation and
floods. Thus, the projections of future floods are also dependent on
people's behaviors.

Compared with the floods, the long-term changes in droughts are more
related to the large-scale and regional-scale climate changes. The
long-term changes in drought have been evaluated, which is closely
related to the past climate changes. Thus, it is quite possible that the
future changes in droughts are also related to future climate changes.
However, it does not intend that the future climate projection is easily
possible.

\begin{figure}[htbp]
\centering
\includegraphics[width=12cm,angle=270]{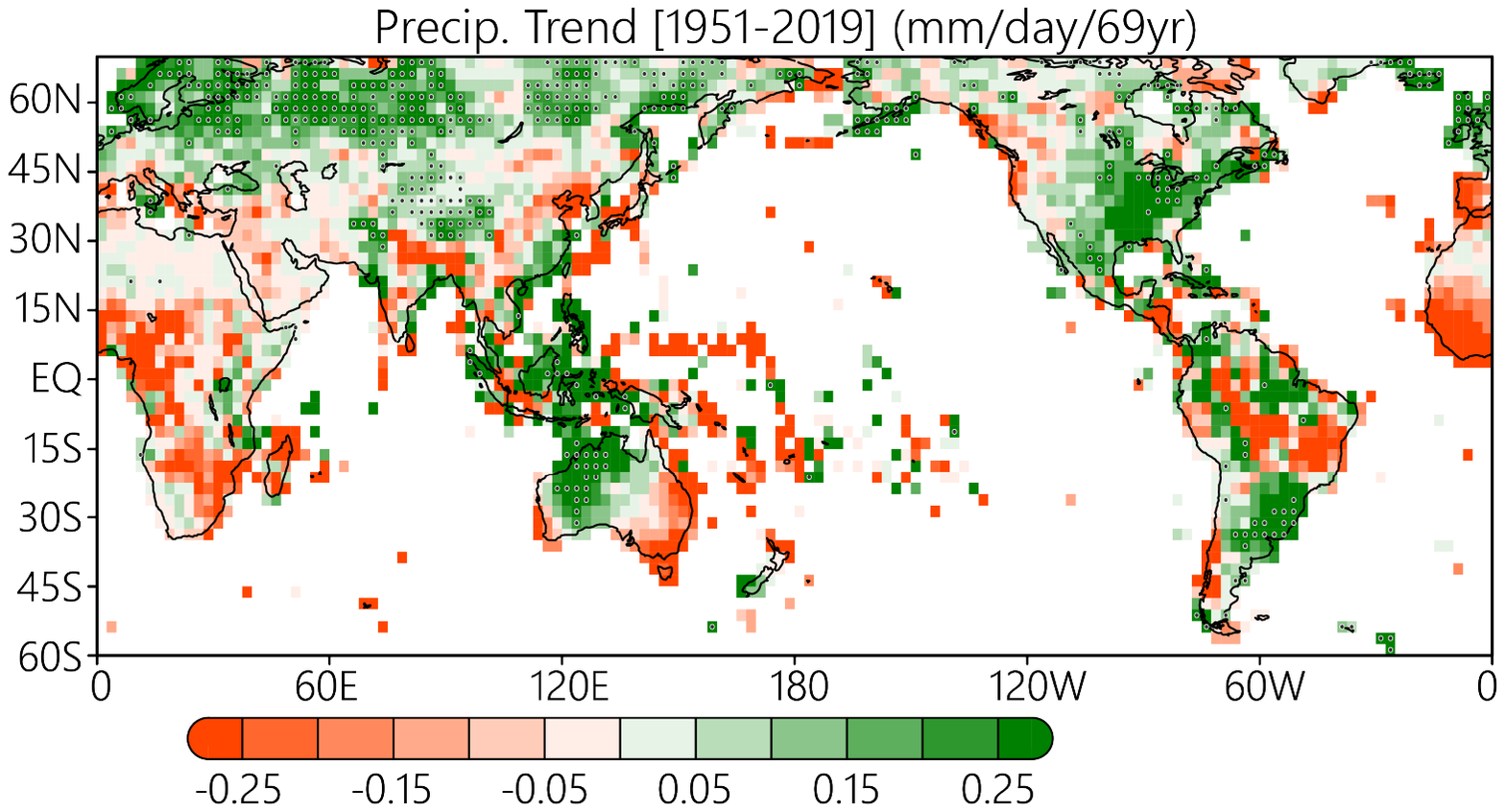}
\caption{Long-term (over 69 years from 1951 to 2019) trends of annual
precipitation over land. Used precipitation data was GPCC (see Section
46.5.1 in this chapter). Spatial and temporal resolution is 2.5$^{\circ}$ $\times$ 2.5$^{\circ}$ and monthly. The unit of the trend is mm/day/69 years. Although values can
be seen over the oceanic regions, this data actively uses precipitation
observations over islands. The trend values are calculated by Sen's
slope estimator, which robustly estimates the trend value with small
outlier adverse effects. The dark grey dots denote that the trend values
are statistically significant at a 95\% significance level, which was
determined by the Mann-Kendall trend test with degrees of freedom being
67.}
\end{figure}

Figure 4 shows the long-term changes in precipitation over the recent
approximately 70 years. The long-term changes in precipitation have a
regional dependency. For example, decreased precipitation was partly
observed in the South, Southeast, East Asian monsoon regions. Over the
African Continent, long-term decreases in precipitation were widely
observed over the near Equator region, which is well-known as one of the
distinct climate changes in the second half of the 20th century. Over
western Europe, the Mediterranean Sea region experienced a long-term
decrease in precipitation, which is associated with severe heatwaves. On
the other hand, precipitation has broadly increased over the northern
region of western Europe. Over the North American continent, the
east-west contrast in the long-term changes is quite significant. The
drying trends over the western parts of North America can be associated
with the recent drought events. A significant decrease in precipitation
can be partly seen over the Amazon region, South America. Over the
Australian Continent, notable east-west gradients of precipitation
trends are observed. The significant drying over the eastern coast of
Australia is closely associated with the recent severe Australian
bushfire (see Fig. 3 of this chapter).

Parts of these long-term decreases in precipitation were associated with
the ENSO variability (e.g., {[}22{]} Dai 2013). The extent to which ENSO
can explain these trends needs to be further explored.

Figure 5 shows an interannual coefficient of variation of annual
precipitation over land. Higher values indicate greater year-to-year
variability over the past 70 years, i.e., areas that tend to have severe
floods, or droughts, or both. Specifically, high values were calculated
over the equatorial pacific and the Maritime Continent, eastern Africa,
the Mediterranean, parts of North and South America, and the Australian
Continent. As for the long-term trends in annual precipitation, ENSO
variability explains the vigorous interannual variations in
precipitation. However, it is quite possible that parts of higher values
imply the recent frequent floods and droughts. A similar analysis has
been applied for the future climate projection (e.g., {[}23{]} Kitoh et
al.~1997; {[}24{]} Kamizawa \& Takahashi 2018), which suggest that
wetting projection on the regional scale is not organized in CMIP5
(Coupled Model Intercomparison Project Phase 5) climate models, whereas
drying projection can have a tendency to occur widely and
systematically. Although there is a random chance in the increasing
trend of floods, and it is challenging to accurately predict the areas
where floods will frequently occur, the increase in droughts is likely
to increase systematically.

\begin{figure}[htbp]
\centering
\includegraphics[width=12cm,angle=270]{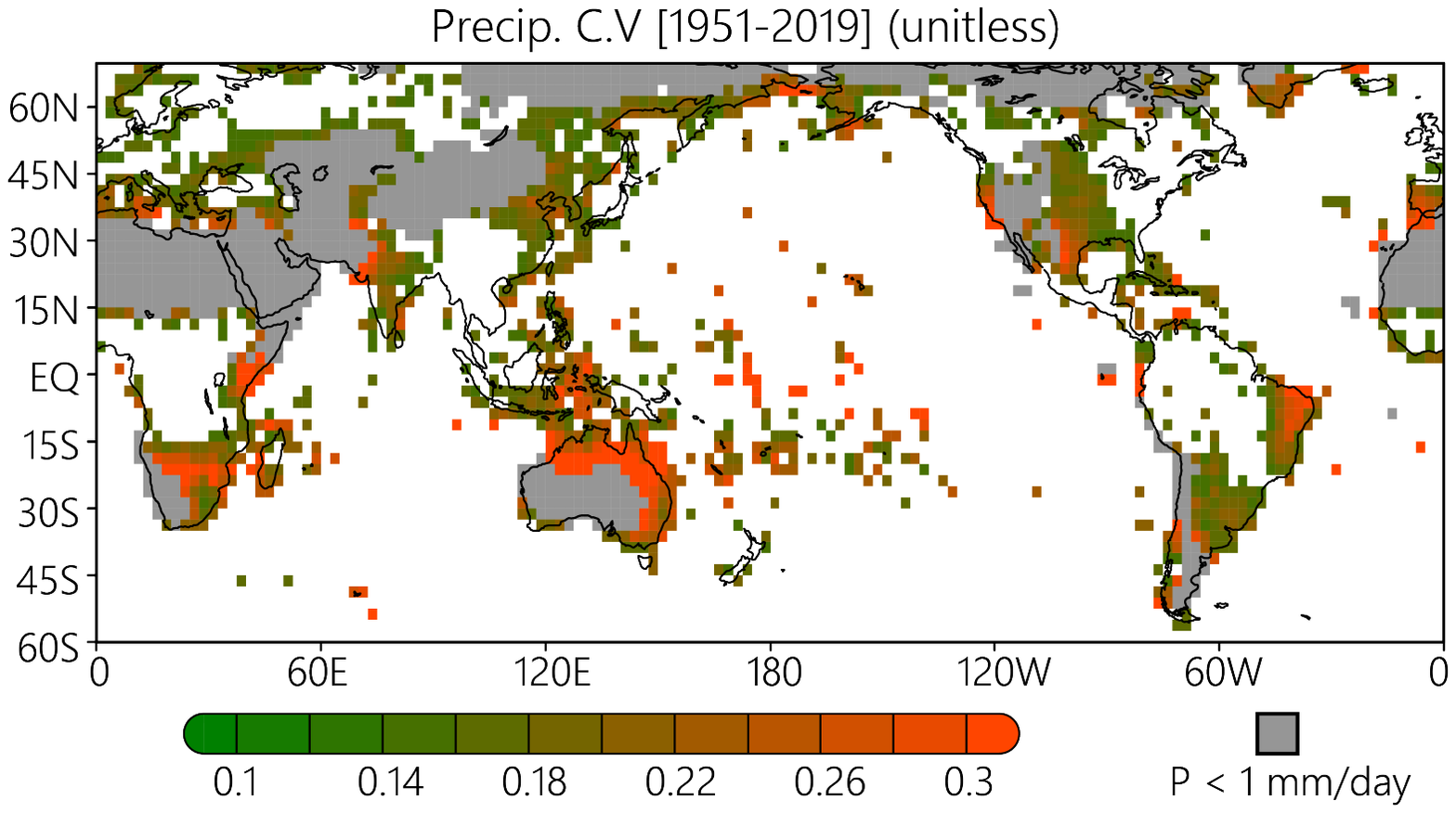}
\caption{Interannual (over 69 years from 1951 to 2019) coefficient of
variation of annual precipitation over land. The coefficient of
variation is the standard deviation divided by the average value, which
indicates normalized standard deviation (dimensionless). A larger value
shows that heavy interannual variability, which implies floods and
droughts frequently occur. The plotted coefficient of variation is
larger than 0.1. The gray shading denotes the annual averaged
precipitation is less than 1 mm/day, which can be a dry climate zone. As
the reference, the global average (including over the ocean)
precipitation is estimated approximately 3 mm/day.}
\end{figure}

It is quite important that air pollution can become seriously related to
the drying tendency of future climate changes. The reduction of the
effect of wet deposition, an increase of the aerosol emission from
natural and artificial biomass burning, and an increase of the aerosol
emission from the dried surface can be easy to imagine. Actually,
possible feedback between surface drying and aerosol loading was
discussed ({[}25{]} Yamaji, and Takahashi 2014).

Thus, as mentioned earlier, it is important to note that even with high
rainfall (low rainfall), there are areas where flooding (drought) occurs
and areas where it does not, as well as timing and time-scales of
occurrence. From the perspective of disaster prevention and future
projection, regional different disaster preventions against floods and
droughts are required because regional problems cannot be solved by
global projection alone.

\subsection{Floods, Drought and Heatwaves in East Asia}

Anomalous precipitation is observed over East Asia when the subtropical
high develops over the northwestern Pacific and abundant moisture
transports along with its anticyclonic circulation. A spatial pattern of
global-scale SST over the Pacific and Indian Oceans controls the
interannual variability in the subtropical high over the northwestern
Pacific (SHNWP), which is well summarized in {[}26{]} Xie et al.~(2016).
For example, an anticyclonic circulation anomaly appears over the
surrounding regions of the Philippines in the El Niño winter, which
suppresses convection there. This anomaly is maintained during the
post-El Niño spring and summer by feedback between SST cooling at its
southeastern flank in association with the enhancement of northeast
trade wind and a reinforcing of the anticyclonic circulation anomaly by
convection suppression over the cool SST. The El Niño also induces the
Indian Ocean warming, which acts to anchor the anticyclonic circulation
anomaly over the western Pacific until summer. As a result, the SHNWP is
intensified in the post-El Niño summer, and an anomalous anticyclonic
circulation transports moisture effectively from the tropical Pacific to
central-eastern China and southwestern Japan. In addition, a cyclonic
circulation anomaly is formed at the north of anticyclonic circulation
anomaly, i.e., over northeastern China and central and northern Japan,
by a poleward propagation of the Rossby wave called as the Pacific-Japan
(PJ) pattern, which activates convection over the Yangtze River basin,
Korea, and Japan. Meanwhile, the genesis of tropical cyclone (TC) is
suppressed over most of the tropical northwestern pacific in the post-El
Niño summer, and the landfall of TC also reduces eastern China and
Korea.

Actually, summer precipitation was extremely high over the Yangtze River
valley in 1983 and 1998, of which summers were in the decaying phase of
strong~El Niño events as well as having the strong warming in the
tropical Indian Ocean. The strong El Niño events were also recorded in
2015/2016, and heavy precipitation was observed in June and July, 2016
over the Yangtze River valley; however, precipitation was less in August
than the normal due to the enhanced mid-latitude teleconnection, which
is referred to as the Silk Road pattern (explained in below). Heavy
precipitation lasted during Meiyu/Baiu season in 2020 over China and
Japan even though the El Niño was not seen in the previous winter.
Although causes of extreme precipitation occurrence in 2020 are
investigated by many researchers as of 2021 spring (e.g., {[}27{]}
Takahashi and Fujinami 2021), the warming in the Indian Ocean associated
with the Indian Ocean Dipole mode is considered as a key factor to make
such a critical situation ({[}28{]} Takaya et al.~2020). In these years,
floods and their huge damages are also recorded. For example, the flood
that occurred in the 1998 summer led to more than 1500 deaths and
economic loss of approximately 255 billion RMB, while the dead and
missing people were less than 150 people and economic loss was estimated
at approximately 120 billion RMB in association with the 2020 flooding
({[}29{]} Wei et al.~2020). Flood disaster is likely to be mitigated by
the infrastructure, ecosystem restoration, progress of weather and
hydrological forecast technology, and so on.

We also should describe the effect of plateau-scale disturbances
generated over the Tibetan Plateau on the heavy precipitation occurrence
in East Asia (e.g., {[}30{]} Sugimoto 2020). The plateau-scale
disturbance is formed during summer by the strong land-surface heating
and by a shear line between moist southwesterly winds over the
southwestern plateau and northeasterly winds associated with
anticyclonic circulation over the northern plateau. A dry-wet gradient
in soil moisture is found from northwest to southeast over the plateau,
and it increases the occurrence frequency of the plateau-scale
disturbances over the western and northwestern plateau. The disturbance
generates the mesoscale convective systems (MCSs) over the southern and
eastern parts of the plateau, where soil moisture is relatively high.
Approximately 25\% of the plateau-scale disturbances with the MCS
propagate eastward and move out of the Tibetan Plateau, and then, they
influence the heavy precipitation occurrence over the upper and middle
areas of the Yangtze River basin. Furthermore, the plateau-scale
disturbance enhances the Meiyu/Baiu precipitation and causes an abundant
moisture inflow toward the Meiyu/Baiu front, which also affects heavy
precipitation events over the lower reaches of the Yangtze River and
southwestern part of Japan.

On the contrary, extreme hot events with convection suppression occur
over East Asia when a high-pressure anomaly appears there.
Climatologically, the western part of SHNWP expands northward and covers
East Asia in August, and the South Asian High (SAH) lies in the upper
troposphere over Eurasia during summer. The center of SAH is found over
the Himalayan regions and oscillates east-west direction with a
time-scale of approximately 10--20 days. The Eastern edge
of the SAH sometimes extends eastward and reaches over the East Asian
countries. The Rossby wave propagation along the Asian jet is a key
factor in developing the SAH and the SHNWP. A major Rossby wave train,
which excited over the Mediterranean and the Aral Sea, is referred to as
the Silk Road pattern ({[}31{]} Enomoto et al.~2003). The Silk Road
pattern intensifies an equivalent-barotropic ridge structure near Japan.
The Rossby wave train induced by atmospheric heating over the Tibetan
Plateau also barotropically enhances an anticyclonic circulation over
the east of Japan as well as the SHNWP intensification associated with
the wave train along with the monsoon westerly in the lower troposphere.
For example, extremely hot summers were recorded over Japan in 1994,
2010, and 2013 under the condition of the development of the SAH and the
SHNWP. A near-surface temperature averaged over Japan was the highest in
2010 since the record began in 1898, and that record is still unbroken
as of spring of 2021. In addition to the continuous hot weather,
precipitation was lower than 50\% of climatology in Japan except for
some regions during the summer of 1994. In this year, water intake from
the rivers was regulated in many areas, and water level in Lake Biwa,
which is the largest lake in Japan, was the lowest on record. A frequent
occurrence of the hot events will also be affected by the global-scale
warming, and heat island effect due to the urbanization and foehn at the
mountain leeward give a local-scale heterogeneity in the spatial
distribution of extreme hot events.

\subsection{Floods in Southeast Asia}

Southeast Asia is consisting of the Eurasian continental portion and
islands area. Compared with the mid-latitude zone, there is not so often
occurred the catastrophic water-related hazards with abundant death
toll, except for storm-surge floods in mainly Philippines.

Based on EM-DAT database since 1900, the deadliest floods occurred by
Cyclone Nargis in early May, 2008. Nargis made landfall in low
Ayeyarwady delta on May 2 with strong wind, consequently the
storm-surge, which height is around 3.5m, claimed at least 138,000 life.
Landfall to Myanmar is very rare, and people has not informed because of
military administration, are considerable reason of abundant death toll.
Second largest case of death toll is Typhoon Haiyan in November, 2013.
Haiyan attacked a central part of Philippine with strong wind (more than
60m/s) and storm surge (at least 5-6m), consequently the death toll is
7,354. Philippine, especially, is prone region for typhoon related
phenomena such as storm-surge, strong wind, and so on. On the other
hand, the death toll is not so much (less than 1,000) in case of
riverine flood due to seasonal abundant rainfall rather than the normal.
However, the period affected due to flood is longer than that by typhoon
or cyclone. Affected period's length directory/indirectly caused in
economic loss. Most catastrophic economic damage is 40 x 109 US\$ in
case of the Chaophraya River flood in central Thailand in 2011. In case
of the 2011 Thailand flood, the largely industrial parks were inundated
about a few months. So even though flood speed is slow thus the death
toll is small, the economic loss could be huge if high property area is
inundated. Flood speed is slow means that the flood water remains a long
time. In Southeast Asia, there are three low-lying area in Chaophraya
River in Thailand, Ayeyarwady River in Myanmar, and Mekong River from
Cambodia to southern Vietnam. Due to sea level rising by climate change,
these areas become more flood-prone region. Recently the urban flood due
to the heavy rainfall in short time is focus from the viewpoint of
economic issues. Although death toll is about 30 in case of urban flood
in Jakarta and Manila, the capital of Indonesia and Philippine, the
economic loss list is second and third economic loss in flood case
excluding storm-surge. Due to recent economic growth, the abundant
properties are accumulated in urban area. When we consider the counter
measures of flood in urban, this accumulation of property become one of
the important issues.

From here, the meteorological conditions of the Thailand severe floods
in 2011 are explained. The meteorological situation of a flash flood due
to a tropical cyclone in tropical regions can be easily understood.
Here, the meteorological condition of the seasonal flood over Southeast
Asia is explained. In 2011, Thailand experienced a severe flood in the
Chao Phraya River Basin. This severe flood was due to the active
tropical disturbance activity ({[}32{]} Takahashi et al., 2015). The
time scale of this flood is seasonal, and the excessively accumulated
rainfall during the monsoon season induced the flood. To investigate the
atmospheric circulations associated with high rainfall, spatial patterns
of rainfall variation were identified. The figure shows the regression
coefficients plotted against the rainfall variability in the region
centered on Thailand, focusing on the severe flooding that occurred in
Thailand in 2011 (Fig. 6).

\begin{figure}[htbp]
\centering
\includegraphics[width=12cm,angle=0]{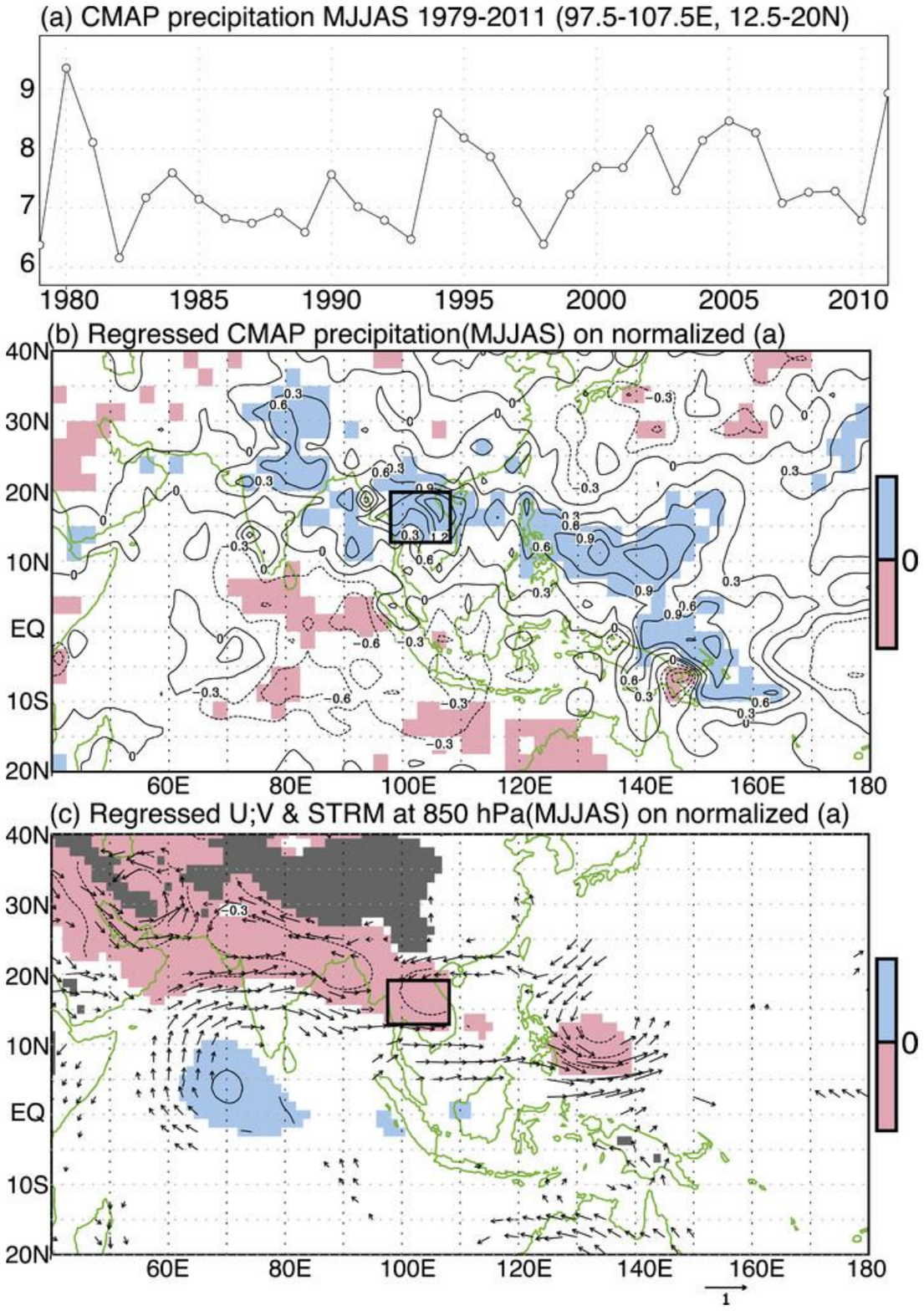}
\caption{(a) Precipitation time series generated from the CMAP dataset for the rainy season (May--September) over the reference region of Indochina (12.5$^{\circ}$--20$^{\circ}$N, 97.5$^{\circ}$--107.5$^{\circ}$E) from 1979--2011. The reference region is used for the regression analysis in (b), (c) and is indicated by a solid rectangle in these panels. (b) Regression of CMAP data during the rainy season against the normalized data (mm day$^{-1}$) shown in (a) from 1979 to 2011. (c) As in (b), but for the 850-hPa zonal and meridional winds and streamfunction (colors) during the rainy season. Areas with colors in (b) and plotted vectors (wind; m s$^{-1}$) and contours and colors (streamfunction; 10$^{6}$ m$^{2}$ s$^{-1}$) in (c) are statistically significant at the 90\% level, as determined by correlation coefficients based on 31 degrees of freedom (df). (Citation: Fig. 5 of {[}32{]} Takahashi \textit{et al.} 2015, copyright 2015 American Meteorological Society)}
\end{figure}

The positive signals were found from northern India, Bangladesh,
Myanmar, Thailand, and further to the Philippines, and there is a
tendency to have more precipitation in similar years. Because the zonal
tiling band corresponded to the significant tropical cyclone path, this
result suggests the importance of tropical disturbance activity for the
excessive rainfall over Thailand and the flood.

\subsection{Drought in Europe}

Over Europe, severe heatwaves or droughts have been significant problems
as an impact of climate change. In 2003 and 2010, the severe heatwaves
were the major meteorological and climatological disasters, which may be
associated with the current climate changes. As the mean state in the
future, severe heatwaves and drought have been projected over and around
the Mediterranean Sea. Most "\textit{Fifth Assessment Report of the
United Nations Intergovernmental Panel on Climate Change (IPCC-AR5)}"
climate models have projected the northward expansion of the Hadley
circulation over Europe in the northern summer, which would induce the
frequent occurrence of heatwaves or more severe droughts. In addition to
the climate changes in the mean state, the interannual variability is
associated with respective heatwave and drought events. If the amplitude
of the interannual variability is the same as the current climate, the
drying trend over Europe enhances respective heatwaves and droughts.

Each heatwave and drought event is associated with the development of a
blocking high over Europe. Usually, the blocking high over Europe lasts
for one to two weeks, which induces dry and hot conditions, particularly
over the inland regions of Europe. However, the location of the blocking
and atmospheric circulation varies by heatwave and drought cases.
Nevertheless, the spatial scale of the heatwaves and droughts are
synoptic rather than regional.

For understanding the heatwaves over Europe, it is essential to consider
the land-atmospheric interactions or feedback during a heatwave event.
When a blocking high forms over Europe, surface air temperature
increases at first. When the surface air temperature is high, more soil
water is consumed to evapotranspiration (summation of soil and water
body evaporation and vegetation transpiration). When the soil has some
available water, the increase in surface air temperature would be
suppressed. However, a long-lasting blocking high can consume much soil
moisture. After the shortage of available soil moisture, most of the
absorbed solar energy at the surface is redistributed to the sensible
heating at the surface, which, in turn, drastically increases the
surface air temperature. Moreover, under the condition of the high
surface air temperature, the surface is provided with more energy due to
an increase in the downward longwave radiation from the atmosphere at
the surface in the surface energy budget relationship. Then, the surface
must consume more energy to increase the surface air temperature. This
land-atmospheric feedback intensifies the heatwaves. Thus, the
atmospheric processes and land-atmospheric interactions play a vital
role in enhancing heatwaves and droughts.

\subsection{Droughts in America}

Over North America, 2010's had severe droughts, such as
California drought or heatwaves. Detailed and more professional review
is provided in a recent review paper ({[}33{]} Seager and Hoerling
2014). The following description refers to this paper. North America has
experienced severe multiyear droughts, such as in 1930's
(called Dust Bowl), in 1950's, 1998 to 2004, and 2010's. Although a seasonal drought has been addressed, which can be explained by the Rossby wave propagation forced by the
anomalous warm SST over the tropical Pacific Ocean (in the El Niño phase), a multiyear drought had not been able to be explained until 2000's.

The break-through study ({[}34{]} Schubert et al.~2004) addressed Dust
Bowl drought in 1930's by a large-ensemble experiment (at that time; at
the time of writing this handbook, experiments with more than a few
dozen ensembles are referred to as large ensemble experiments, but this
term may change in the past, present, or future) using an atmospheric
general circulation model (AGCM) prescribed the observed SST. A
large-ensemble experiment enables us to evaluate the contribution of
oceanic impact (SST anomaly) and internal atmospheric and land
conditions variability, for example, which can explain that a phenomenon
is associated with the land-atmospheric feedback (see Section 4.4 in
this chapter). For the multiyear drought, both persistent cold tropical
Pacific and warm tropical North Atlantic SST anomalies are the drivers.

As explained in Section 4.4 in this chapter, an external forcing is
necessary to escape from the loop of droughts, such as the multiyear
drought. Over North America, wet and warm low-level southerlies from the
Gulf of Mexico, which are referred to as the Great Plain low-level jet,
bring the abundant moisture from the tropical region. The wet low-level
southerlies are associated with abundant precipitation, which can
recover the normal climate situation.

Not only the causes of the drought and heatwaves, but the anthropogenic
impacts have also been detected by the Event Attribution method. A
previous study on the California drought indicated that anthropogenic
warming has substantially increased the overall likelihood of extreme
California droughts ({[}35{]} Willams et al.~2015).

\section{Summary}

This section addresses the flood and drought over the world under the
changing climate. As shown above, the causes and temporal and spatial
scales are quite regional and seasonal dependent over the world. Thus,
an important, though not concrete, the summary is that it is very
difficult to discuss future floods and droughts through a unified set of
indicators and methods. Even if uniform indicators and methods are used,
specific assessments and recommendations need to be made, taking into
account regional and seasonal differences in the interpretation of the
indicators and methods.

In the context of global warming, floods over some regions are
associated with short-time heavy precipitation events. On the other
hand, over other regions, precipitation of a future super tropical
cyclone may affect floods.

Drought is sometimes associated with the long-term changes in drying
trends due to an expansion of the Hadley circulation, such as the
Mediterranean Sea region. On the other hand, decadal or interannual
variations in the atmosphere-ocean coupling modes, which are understood
as internal climate variations, can control severe drought in some
regions.

It should be noted that a specific flood/drought event can be explained
by the specific precipitation and the related specific atmospheric
circulations. Also, a specific season of extreme flood/drought can be
explained by anomalous precipitation and the related atmospheric
circulations. For each event, it is difficult to separate climate change
from natural variability, but a number of studies have found that the
frequency of floods and droughts can vary with climate change. Thus,
each specific flood/drought is partly, but certainly, affected by
climate changes.

\end{document}